\documentclass[prl,twocolumn,showpacs,preprintnumbers,amsmath,amssymb,superscriptaddress, bibnotes, dvipdfmx]{revtex4}

\usepackage{color}
\usepackage{graphicx}
\usepackage{dcolumn}
\usepackage{bm}
\usepackage{mathrsfs}

\begin{document}

\title{Spin Polarization Enhanced by Spin-Triplet Pairing in Sr$_2$RuO$_4$ Probed by NMR}

\author{K.~Ishida}
\email{kishida@scphys.kyoto-u.ac.jp}
\author{M.~Manago}
\author{T.~Yamanaka}
\author{H.~Fukazawa}
\altaffiliation{Present Address: Department of Physics, Chiba University, Japan}
\author{Z.~Q.~Mao}
\altaffiliation{Present address: Department of Physics, Tulane University, USA }
\author{Y.~Maeno}
\affiliation{Department of Physics, Graduate School of Science, Kyoto University, Kyoto 606-8502, Japan}
\author{K.~Miyake}
\affiliation{Toyota Physical and Chemical Research Institute, Nagakute, Aichi 480-1192, Japan }

\date{\today}

\begin{abstract}
We report a novel phenomenon intimately related to the spin-triplet superconductivity. 
It is well known that the spin susceptibility decreases below the superconducting transition temperature in almost all superconductors because of spin-singlet pair formation, while it may remain unchanged in a handful of spin-triplet exceptions. 
Here we report the observation in Sr$_2$RuO$_4$ with nuclear magnetic resonance (NMR) that the spin susceptibility originating from the Ru-4$d$ electron slightly $increases$ by $\sim 2 $\% of total and becomes inhomogeneous in the superconducting state. 
These are reasonably explained if the electron pairs form the equal-spin-pairing (ESP) in the mixed state. 
A similar phenomenon was predicted for superfluid $^3$He forty years ago, but had never been demonstrated in any superconductor. 
\end{abstract}


\abovecaptionskip=-5pt
\belowcaptionskip=-10pt

\maketitle

Sr$_2$RuO$_4$ (SRO) with the same tetragonal structure as the high-temperature superconductor La$_{2-x}$Sr$_x$CuO$_4$ shows the unconventional superconductivity at 1.5 K on the ground of a well-characterized Fermi-liquid state\cite{Maeno94Nature}. 
This implies that the superconductivity in SRO can be interpreted based on the weak-coupling theory more generally.
To understand the superconducting (SC) nature, one of the most important properties of superconductivity is the spin state of the SC pairs. 
We have measured the Knight shift at the $^{17}$O and $^{99}$Ru sites of SRO, particularly in the SC state, since the measurement of the Knight shift probing the static hyperfine field at the nuclear site is one of the most reliable methods to measure the spin susceptibility \cite{MacLaughlin76}. 
We have reported that the Knight shifts of $^{17}$O \cite{Ishida98Nature, Mukuda99JLTP} and $^{99}$Ru \cite{Ishida01PRB,Murakawa04PRL} nuclei are unchanged in the SC state. 
The unchanged spin susceptibility, which was supported also from the polarized neutron scattering measurements \cite{Duffy00PRL}, strongly suggests that the equal-spin-pairing (ESP) state of the spin-triplet superconductivity is realized in SRO, in which the SC pairs consist of the up-up ($|\uparrow \uparrow \rangle$) or down-down ($|\downarrow \downarrow \rangle$) pairs.  
In addition, the $\mu$SR \cite{Luke98Nature} and Kerr-effect \cite{Xia06PRL} measurements in the SC state suggest the broken time-reversal symmetry. 
Unlike heavy-fermion systems, the effect of spin-orbit coupling should be relatively weak, the pairing state in SRO would be closely analogous to the case of superfluid $^3$He.
By taking the experimental results into account, the chiral $p$-wave spin-triplet state, which is analogous to the $A$-phase of superfluid $^3$He \cite{Leggett75RMP}, has been considered as the most promising candidate for the SC pairing state \cite{Mackenzie03RMP,Maeno12JPSJ}. 

However, toward the establishment of this pairing state, there still remains a controversy since the superconductivity is strongly suppressed with a first-order transition under the in-plane magnetic fields near the SC critical field $H_{c2}$ as shown in Fig.~1 (a) \cite{Kittaka09PRB,Yonezawa13PRL}.
In addition, sharp magnetization jump with the hysteresis at $H_{c2}$ was observed when the magnetic field is exactly parallel to the $ab$ plane at 0.1 K \cite{Kittaka14PRB}. 
These are phenomena usually expected in a spin-singlet superconductor \cite{Clogstone62PRL,Machida08PRB}. 
Furthermore, non-detection of some of the behaviors expected for the chiral $p$-wave state (e.g. chiral edge current \cite{Matsumoto99JPSJ, Bjorsson05PRB, Kirtley07PRB} and splitting of $T_c$ by in-plane magnetic fields of any magnitude \cite{Deguchi02JPSJ}) casts some doubts on this pairing state. 
Therefore, convincing evidence for establishing the SC pairing state in SRO, particularly finding a new phenomenon specific to the pairing state, has been desired. 

In this paper, we report, from the ``double-site'' Knight-shift measurement in the SC state, which is the reliable method to subtract the Meissner effect from the observed Knight shift, that the spin susceptibility of SRO becomes inhomogeneous and its average slightly {\it increases} in the SC state. 
The small increase of spin susceptibility cannot be explained with a singlet-pairing state but consistently interpreted with the ESP state in triplet superconductivity. 
As far as we know, the increase of the spin-susceptibility in the SC state has not been reported in any superconductors, thus would be a new phenomenon specific to spin-triplet superconductors.
\begin{figure}[tb]
\begin{center}
\includegraphics[width=8.7cm,clip]{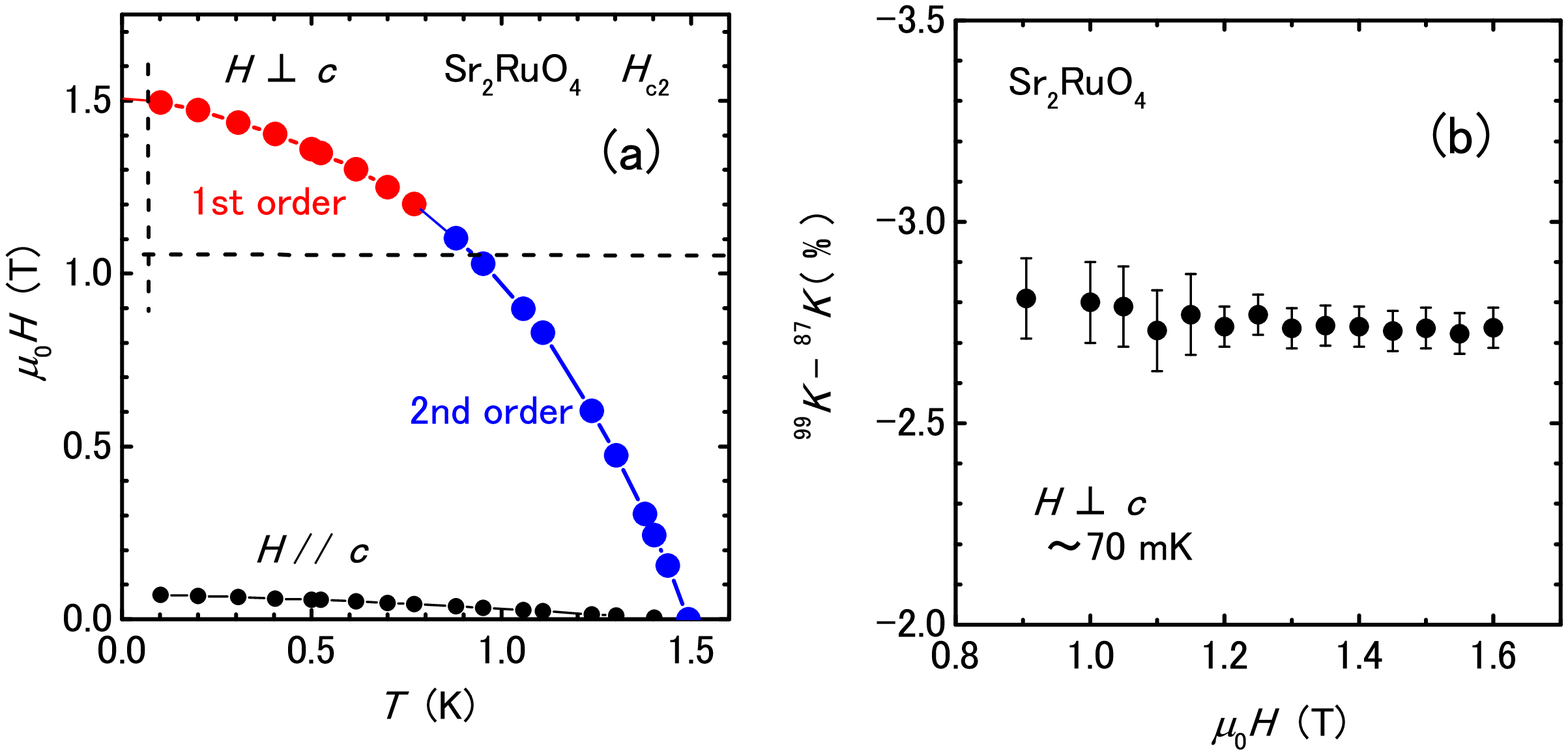}
\end{center}
\caption{(Color online) SC phase diagram of the present SRO sample determined by $^{ac}\chi$ in $H$ parallel and perpendicular to the $c$ axis. The $H$ region where the SC transition is of the 1st [2nd] order is shown by the red [blue] points and lines. The vertical [horizontal] dotted arrow shows the $H $($T$) scan measurement shown in Fig.~2 (a) [Fig.~3 (a)]. Field dependence of the difference in the Knight shift between $^{99}$Ru ($^{99}K$) and $^{87}$Sr ($^{87}K$) measured at around 70 mK. }
\label{Fig.1}
\end{figure}

We performed $^{99}$Ru- and $^{87}$Sr-NMR on single crystals of SRO ($T_c \sim$ 1.5 K) under a precise control of the angle between the $a$ axis and external fields with the accuracy of better than 0.5$^{\circ}$. 
Low-temperature NMR and Meissner measurements were carried out with a $^3$He-$^4$He dilution refrigerator, in which the single-crystalline SRO is immersed into the $^3$He-$^4$He mixture to avoid RF heating during the measurements. 
In addition, we measured $^{99}$Ru and $^{87}$Sr Knight shifts, at the same field and pulse conditions\cite{supplemental1}. 
The Knight shift $K$ at a constant external magnetic field is defined as the frequency shift from a reference frequency ($f_{\rm ref}$), $f - f_{\rm ref}$, divided by $f_{\rm ref}$, i.e., $K \equiv (f / f_{\rm ref}) -1$. 
The double-site Knight-shift measurement is the Knight-shift measurements on two different nuclei in the same compound, which is particularly useful for the precise extraction of spin-susceptibility in the SC state. 
We denote the spin and orbital shifts at the $i$ [$i$ = 99(Ru) and 87(Sr)] site as $^{i}K_{s}(T,H)$ and $^{i}K_{\rm orb}$, respectively; $^{i}K_{\rm orb}$ should be a positive value and is usually temperature and field independent.  
The observed total NMR shift $^{i}K(T,H)$ at the $i$ site in the SC state is expressed by  
\begin{eqnarray*}
\lefteqn{^{i}K(T,H)}\\
 & = & \left[1+^{i}K_s(T,H)+^{i}K_{\rm orb} \right]\left[1+K_{\rm dia}(T,H) \right] - 1\\
 & \approx & ^{i}K_s(T,H) + ^{i}K_{\rm orb} + K_{\rm dia}(T,H)  
\end{eqnarray*}
The factor $1+K_{\rm dia}$ takes into account of the SC diamagnetic reduction of the magnetic field due to the Meissner effect. 
$K_{\rm dia}(T, H)$ is denoted as $\Delta B(T)/H$ with SC diamagnetization $\Delta B(T)$ and is the same for all nuclear species in the sample because it is a bulk  phenomenon ascribed to the SC screening current varying in the scale of the London penetration depth [$\lambda_{//ab} \approx 0.15 \mu$m, $\lambda_{//c} \approx 3 \mu$m at $T \rightarrow 0$] \cite{Maeno12JPSJ, Akima99JPSJ}. 
In addition, higher order terms are neglected because $K_{\rm dia}$ is four orders of magnitude smaller than unity. 
Thus the difference between $^{99}K(T, H)$ and $^{87}K(T,H)$ does not contain the $K_{\rm dia}(T, H)$ term and is expressed as, 
\begin{eqnarray*}
\lefteqn{^{99}K(T,H)-^{87}K(T,H)}\\
 & = & \left[^{99}K_s(T,H)-^{87}K_s(T,H)\right]+^{99}K_{\rm orb}-^{87}K_{\rm orb}\\
 & = & \left(^{99}A_{\rm hf}-^{87}A_{\rm hf} \right)\chi_s(T,H) + {\rm const.},  
\end{eqnarray*}                   
where $^{i}A_{\rm hf}$ is the hyperfine coupling field at the $i$ site and $\chi_{s}(T, H)$ is the spin susceptibility per a Ru atom and the relation of $^{i}K_s(T, H) =^{i}A_{\rm hf} \chi_{s}(T, H)$ is used. 
It is crucially important to note that the $^{99}A_{\rm hf}$ at $^{99}$Ru is approximately two-orders of magnitude larger than $^{87}A_{\rm hf}$ at $^{87}$Sr, ($^{99}A_{\rm hf} \sim  -25$ T/$\mu_{\rm B}$ \cite{Ishida01PRB} and $^{87}A_{\rm hf} \sim 0.3$ T/$\mu_{\rm B}$ \cite{Ishida03PRB} ). 
It should also be noted here that the sign of $^{99}A_{\rm hf}$ is negative due to the core polarization effect by Ru-4$d$ electrons in contrast with small and positive $^{87}A_{\rm hf}$ due to Sr-5$s$ electrons. 
Roughly speaking, $^{87}K$ mainly detects the interior field distribution originating from the Meisser effect only, whereas $^{99}K$ additionally detects the spin-susceptibility governed by the Ru-4$d$ electrons forming Cooper pairs in the SC state. 

Figure 1 (b) shows the field dependence of the difference $^{99}K -^{87}K$, measured at around 70 mK in the field range between 1.6 and 0.905 T. 
No appreciable change of $^{99}K-^{87}K$ was detected within the accuracy of 0.1 \% down to 1.1 T, although superconductivity sets in below 1.5 T. 
Since the quantity $^{99}K-^{87}K$ is free from the Meissner effect, the invariant $^{99}K-^{87}K$ clearly indicates that the spin-susceptibility in the SC state is unchanged from that in the normal state. 
Interestingly, although the experimental error becomes larger due to the weaker NMR intensity in smaller fields, $^{99}K-^{87}K$ slightly {\it increases} below 1.1 T. 
This tendency is appreciable from the comparison between the NMR spectra in the normal state and those in the SC state at the $^{99}$Ru and $^{87}$Sr sites, which are shown in Fig.~2(a) and Fig.~2(b). 

\begin{figure}[tb]
\begin{center}
\includegraphics[width=8.5cm,clip]{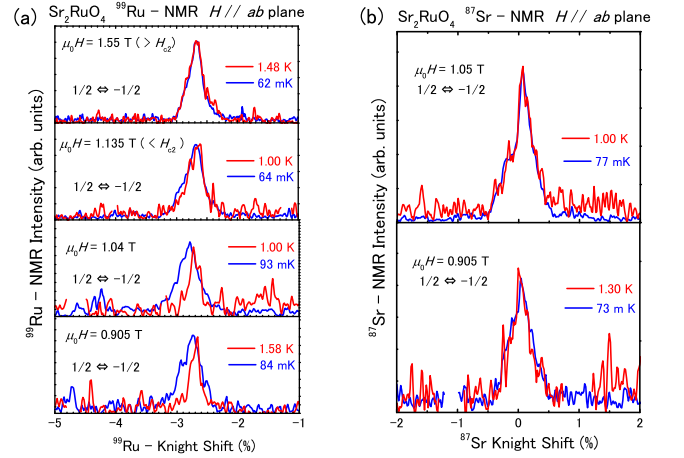}
\end{center}
\caption{(Color online)  (a) $^{99}$Ru-NMR spectra, arising from the central ($-1/2 \leftrightarrow +1/2$) transition, measured in the normal and SC states under the various fields from 1.55 T greater than the in-plane $H_{c2}$ value to 0.905 T below $H_{c2}$. The peak position and linewidth of the NMR spectra at 1.55 T are unchanged, indicating that the electronic state in the normal state is invariant at low temperatures. However, the peak positions in the SC spectra start to decrease below 1.15 T. (b) $^{87}$Sr-NMR signal, arising from the central ($-1/2 \leftrightarrow +1/2$) transition, measured in the normal and SC states under $\mu_0H$ = 1.05 and 0.905 T smaller than $H_{c2}$. The peak position and linewidth of the NMR spectra are almost unchanged, indicating that the change of $^{87}K$ perpendicular to the $c$ axis in the SC state are negligibly small. NMR spectra are normalized with the peak height, and spike noise is erased for eye appeal. }
\label{Fig.2}
\end{figure}

The nearly unchanged $^{87}$Sr-NMR spectra in the SC state are in good agreement with the experimental facts that the SC diamagnetic field is negligibly small due to the plate-like shape of the crystal, estimated as approximately only 0.1- 0.2 mT for the in-plane magnetic fields\cite{Mackenzie03RMP,Rastovski13PRL} ,and that spin susceptibility is small at the $^{87}$Sr site. 
In contrast, the $^{99}$Ru-NMR spectrum, which is totally unchanged under the field of 1.55 T, becomes broader and its ``center of gravity" is slightly shifted to the negative side in the SC state in the fields lower than 1.1 T. 
It should be noted that the negative shift of the $^{99}K$ indicates that the spin-susceptibility increases in the SC state, since the hyperfine coupling constant $^{99}A_{\rm hf}$ at $^{99}$Ru is negative. 
The broadening of the $^{99}$Ru-NMR spectrum seems to be related with the character of the SC transition: the spectrum is almost unchanged in the field range where the SC transition is of the 1st order, but starts to broaden and slightly shifts to the lower frequency when the SC transition is of the second order in the field. 
Such field dependence is attributed to the fact that the SC fraction at low temperatures becomes larger in lower fields.  

The change of the $^{99}$Ru-NMR spectra in the SC state is clearly observed from the $T$-scan measurement at 1.04 T. 
Figure 3 shows (a) $^{99}$Ru NMR spectra arising from the central ($1/2 \leftrightarrow -1/2$) transitions at various temperatures, and the temperature dependencies of (b) Meissner signal, (c) full widths at half maximum (FWHM) of the NMR peaks, and (d) the frequency shift from the normal-state values above $T_c$ ($\delta f$) in both $^{99}$Ru- and $^{87}$Sr-NMR spectra. 
The $^{99}$Ru-NMR spectra become broader in the SC state.
The FWHM and the peak frequency were determined by fitting the observed spectra to the Gaussian function shown in Fig.~3(a).
The FWHM increases gradually below $T_c(H)$, and $\delta f$ appreciably shifts to the negative frequency below 0.5 K.  
 
While such broadening and tiny shift of the spectra ($\Delta$FWHM = $3 \pm 0.5$ kHz, $\delta f = 1.8 \pm 0.5$ kHz) were observed in the previous measurements \cite{Ishida01PRB}, we could not conclude that these behaviors originate from the Ru-4$d$ electrons without the present firm evaluation of the Meissner effect from the $^{87}$Sr-NMR measurement. 
Although the observed change of the $^{99}$Ru-NMR spectrum at 1.04 T is small, it corresponds to the $\sim 1.5$ mT broadening and $\sim 1$ mT shift, which are much larger than the Meissner effect evaluated from the $^{17}$O-NMR \cite{Ishida98Nature} and small-angle neutron scattering measurements \cite{Rastovski13PRL}. 
Therefore, the broadening of the spectra and negative shift of the $^{99}K$ imply that the hyperfine field at $^{99}$Ru, which is proportional to the spin susceptibility produced by the Ru-4$d$ electrons, becomes inhomogeneous and its average slightly increases in the SC state. 

To the best of our knowledge, the increase of the spin susceptibility in the SC state has never been reported so far. 
Particularly, it seems impossible to explain the observed increase of the spin susceptibility at lower temperatures on the basis of a singlet-pairing scenario, since the spin susceptibility of the singlet pairing decreases to zero at $T = 0$ in principle. 
The Knight-shift difference between the normal state ($K_{\rm S}$) and the SC state ($K_{\rm S}$) $\delta K = K_{\rm S} - K_{\rm N}$ of the $^{99}$Ru NMR, estimated to be $\delta ^{99}K \sim  -0.09$ \% at $T \rightarrow  0$, is entirely dominated by the change of the spin part, since the shift by the Meissner effect is negligibly small. 
This gives the increase of the electronic magnetization $\delta m \sim  2.7 \times 10^{-5} \mu_{\rm B}$ in the SC state, which is estimated from the relation $\delta m = (\delta K \mu_0H) /^{99}A_{\rm hf}$, and thus the ratio between $\delta m$ and the normal-state magnetization $m_n$defined by $m_n = \chi_n \mu_{0} H $ is $\delta m/m_n \sim 2.0 \times 10^{-2}$. 
\begin{figure}[tb]
\begin{center}
\includegraphics[width=9cm,clip]{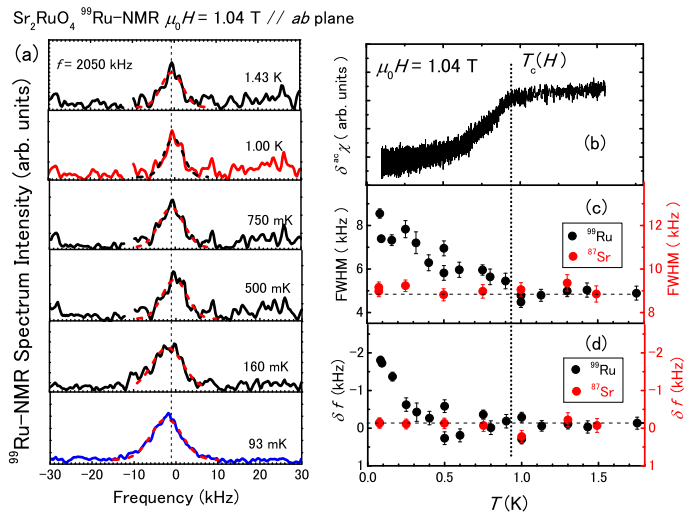}
\end{center}
\caption{(Color online)(a) $^{99}$Ru NMR spectra arising from the central ($1/2 \leftrightarrow -1/2$) transitions measured at various temperatures at 1.04 T. Spike noises are erased for eye appeal. The dotted Gaussian peaks are fittings for the estimation of the FWHM and $\delta f$. Temperature dependence of (b) Meissner signal (c) the full width at half maximum (FWHM) and (d) the frequency shift from the normal-state values above $T_c$ in both $^{99}$Ru- and $^{87}$Sr-NMR signals. The $^{87}$Sr-NMR data were obtained at 1.05 T. The scales of (c) and (d) for $^{99}$Ru ($^{87}$Sr) are shown in the left (right) axes.    }
\label{Fig.3}
\end{figure}

Now, we discuss the origin of our observation in the SC state. 
The magnitude of the spectra broadening is as expected from the field inhomogeneity in the mixed state where the nearly half of the region is approximately regarded as the field-induced normal state.
The question is the origin of the small increase $\delta m$.
One possibility is that $\delta m$ emerges coherently from  a net dipole field by other Cooper-pair spins perpendicular to the Cooper-pair orbital moment, which is well known in the $A$-phase of superfluid $^{3}$He \cite{Leggett75RMP}. 
While this possibility is plausible, this interaction may be negligible after averaged in the present MHz time scale, since the interaction would work in the electronic-spin-resonance frequency range (GHz). 
Thus this effect does not induce any spectral shift nor broadening. 
The other possibility for an additional spin polarization below $T_c$ is due to the spin-orbit (SO) interaction with Cooper-pair orbital momentum along the $c$-axis \cite{Miyake10JPSJ}. 
However, in such a non-unitary state, the additional spin polarization would be along the $c$-axis, and thus is not relevant to the present observation. 
Rather, we point out that the extra magnetization in the SC state is a natural consequence of the ESP state formed by the strongly correlated electrons as discussed in the recent theoretical paper \cite{Miyake14JPSJ}. 
We will next examine this possibility.

In both the normal state and the ESP states under the magnetic field, the density of states (DOS) of the down-spin electrons is larger than those of the up-spin electrons due to the spin polarization. 
Under such difference in DOS, the SC condensation energy of each spin species in the ESP state is different in general. 
This may lead to an additional spin polarization in the SC state, and this polarization is enhanced in the strongly correlated electron system; to gain the free energy associated with the SC condensation, the up-spin and down-spin pairs are redistributed with each other. 
Based on this mechanism, the ratio of the extra magnetization $\delta m$ to the normal magnetization $m_n$ (at $T$ = 0 K) is calculated as \cite{Miyake14JPSJ},
\begin{equation}
\frac{\delta m}{m_n} \approx \frac{\alpha \Delta^2}{4\mu_{\rm B} \epsilon_{\rm F} \mu_0 H}\left(1+\frac{2}{VN_{\rm F}}\right)\left(1+F_0^a\right), 
\end{equation} 
where $\Delta$, $\epsilon_{\rm F}$, $(VN_{\rm F})^{-1}$, and $F_0^a$ are the SC gap, the effective Fermi energy, the SC coupling constant of SRO, and the Landau parameter for correction of the susceptibility, respectively. 
Here, $\alpha$ is a coefficient characterizing the steepness of the slope of the energy-dependent DOS around $\epsilon_{\rm F}$; the energy dependence of the DOS is expressed as,
\begin{equation*}
N(\xi) \cong N_{\rm F}\left(1+\frac{\alpha}{\epsilon_{\rm F}}\xi\right),
\end{equation*}
where $\xi$ is the quasiparticle energy measured from the chemical potential. 
Details of the derivation of eq. (1) are shown in supplemental material \cite{supplemental2}.

The increase of the magnetization induced by this effect was first predicted by S. Takagi as a possible effect in the $A_1$ phase of superfluid $^3$He \cite{Takagi74PTP}. 
We point out that the extra magnetization by this effect is enhanced in SRO since the energy dependence of the DOS near $\epsilon_{\rm F}$ is sharper due to the van-Hove singularity of the $\gamma$-band and the ratio of the SC-gap energy to $\epsilon_{\rm F}$ is larger than that of conventional superconductors. 
In addition, the extra magnetization would be inhomogeneous, since the measurement is done in the SC mixed state, where the normal-state component is induced by the external field.  
By using the physical parameters reported in SRO ($\epsilon_{\rm F} \sim 3.8 \times 10^3$ K, $\Delta \sim 1.7 \times T_c \sim 2.6$ K, $(VN_{\rm F})^{-1} \sim 7$, and $(1+F_0^a) \sim 1/2$ \cite{Miyake10JPSJ}), the enhancement of the magnetization $\delta m/m_n$ is estimated to be  
\begin{equation*}
\frac{\delta m}{m_n} \approx 5.5 \times 10^{-3} \alpha.
\end{equation*}
Although the parameter $\alpha$ in ideal two dimensional Fermi surfaces is known as $\alpha = 0$, the parameter $\alpha$ of the $\gamma$ band, which is active for superconductivity, is estimated to be $\alpha \sim 4$, since the $\gamma$ band is close to the van Hove singularity  \cite{Bergemann03AdvPhys}, which is actually suggested by La-doped Sr$_2$RuO$_4$ \cite{Kikugawa04PRB}.  
Therefore, the experimental finding of $\delta m/m_n \sim 2.0 \times 10^{-2}$ is consistently interpreted with this mechanism.

It should be noted that the extra magnetization $\delta m$ is independent of the magnetic field $H$, which implies that the spontaneous magnetization exists or the time reversal symmetry is spontaneously broken. 
However, this spontaneous magnetization would not be related with the spontaneous field (0.06 mT) detected by the $\mu$SR measurement \cite{Luke98Nature}. 
This is because the effective field probed by muon, which is produced by the spontaneous magnetization, is one-order of magnitude smaller than 0.06 mT due to the smaller hyperfine coupling between the muon and electron spins  \cite{Miyake14JPSJ}. 
On the other hand, the spontaneous field 0.06 mT cannot be detected by the $^{99}$Ru and $^{87}$Sr NMR spectra, since it corresponds to the 0.1 kHz shift due to the small gyromagnetic ratio of the $^{99}$Ru and $^{87}$Sr nuclei. 
This implies that the spontaneous field probed by muon does not originate from the spin component, but from the orbital component of the SC pairs.

Next, we comment on a $T_c$ splitting in the ESP state under magnetic fields. 
It is noted that the same spin polarization would induce a $T_c$ splitting just like the $A_1$ phase of superfuid $^3$He. 
By using the same notation as above, the splitting of $T_c$ in $\mu_0 H = 0.9$ T is estimated as 
\begin{equation*}
\frac{\delta T_c}{T_c} \approx \frac{2\alpha}{VN_{\rm F}}\frac{\mu_{\rm B}\mu_0 H}{\epsilon_{\rm F}} \sim 2.2 \times 10^{-3} \alpha.
\end{equation*}
By taking $\alpha \sim 4$, which is satisfied with the extra magnetization reported here, $\delta T_c$ is $\sim 9$ mK. 
This splitting is smaller than the present experimental resolution, but in principle detectable by other physical quantities such as specific heat.

In conclusion, we found from the double-site Knight-shift measurements that the magnetization originating from the Ru-4$d$ electrons becomes inhomogeneous and its average slightly increases in the SC state, particularly below 1.1 T. 
We show that the extra magnetization is consistently understandable as a characteristic feature of the ESP state, and is a new phenomenon specific to the ESP in triplet superconductivity. 
The present result strongly suggests that the origin of the first-order transition is not ascribed to the Pauli-depairing effect, but to other effects including orbital degree of freedom or new effects which have never been discussed before.

The authors thank H. Murakawa, H. Mukuda, Y. Kitaoka, K. Asayama, S. Yonezawa, H. Ikeda, K. Machida, I. Mazin, V. Mineev, A. Mackenzie, Y. Sasaki, and T. Mizusaki for valuable discussions. This work was partially supported by Kyoto Univ. LTM center, the ``Heavy Electrons'' (No. 20102006) and ``Topological Quantum Phenomena'' (No. 22103002) Grant-in-Aid for Scientific Research on Innovative Areas from the Ministry of Education, Culture, Sports, Science, and Technology (MEXT) of Japan, and  by Grant-in-Aids for Scientific Research (No. 25400369, No. 23244075 and No. 15H05745)  from the Japan Society for the Promotion of Science (JSPS).

\newpage
\section{Supplementary material for `` Spin Polarization Enhanced by Spin-Triplet Pairing in Sr$_2$RuO$_4$ Probed by NMR''}
\

\abovecaptionskip=-5pt
\belowcaptionskip=-10pt

\maketitle

\section{Experimental Details} 
We used three pieces of single crystals for the present measurements. 
The largest crystal was $4 \times 11 \times 1$ mm$^3$. 
These samples were identical to those used in the previous measurements\cite{Ishida01PRB,Murakawa04PRL,Murakawa07JPSJ}
The precise alignment for applied in-plane fields to SRO is achieved from the angle-dependence of the Meissner signal shown in Fig.~1. 
When fields are exactly parallel to the basal plane, the Meissner signal shows a local maximum, which is similar to a Josephson vortex lock-in behavior where the vortices are mobile along the layer. 
This is because the SC anisotropy of SRO is large ($H_{c2,ab}/H_{c2,c} \sim 20$)\cite{Akima99JPSJ}, although the superconductivity in SRO is three-dimensional. 
NMR measurements were carried out under the precise alignment of in-plane field within a 0.5$^{\circ}$.

\begin{figure}[b]
\begin{center}
\includegraphics[width=7cm,clip]{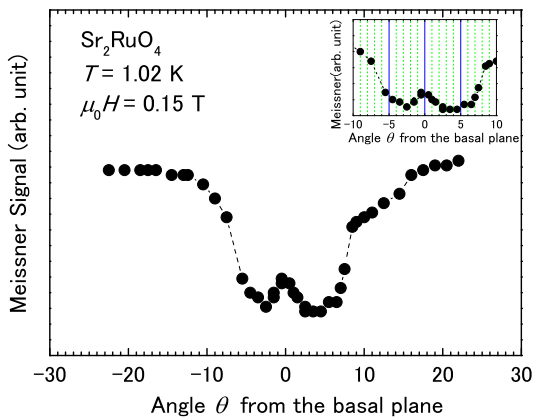}
\end{center}
\caption{(Color online) The dependence of the Meissner signal on the angle between the applied magnetic field and the RuO$_2$ plane measured at 1.02 K under 0.15 T. The inset shows an enlarged view around $\theta = 0$. }
\label{Fig.1}
\end{figure}

Some properties of Sr and Ru isotopes relevant to the present NMR measurements are listed in Table 1. 
NMR frequency of $^{87}$Sr ($I$ = 9/2) and $^{99,101}$Ru ($I$ = 5/2) in SRO under the field along the RuO$_2$ plane $H$ ($x$ direction) can be calculated by solving the secular equations of  
\begin{eqnarray*}
\mathscr{H} &=&\mathscr{H}_Z+\mathscr{H}_Q \\
 &=&-\frac{\gamma_n}{2\pi}h(1+^iK_{ab})I_x \cdot \mu_0H+\frac{h\nu_Q}{6}\left[3I_z^2+I(I+1)\right]
\end{eqnarray*}
where $\mathscr{H}_Z$ and $\mathscr{H}_Q$ are the Zeeman and electric quadrupole interactions. 
$^{i}K_{ab}$ and $^{i}\nu_Q$ are the Knight shift along the RuO$_2$ plane and the electric quadrupole frequency along the principal axis ($c$ axis, $z$ direction) at the $i$ [$i$ = 99(Ru) 101(Ru) and 87(Sr)] site, and these values in SRO were already evaluated from the previous measurements \cite{Murakawa04PRL,Murakawa07JPSJ,Ishida03PRB}. 
The resonance frequencies ($f_{\rm res}$) of $^{87}$Sr, $^{99}$Ru and $^{101}$Ru in SRO under certain magnetic fields parallel to the basal plane are evaluated from the difference between the relevant eigenvalues of $\mathscr{H}$, which are shown in Fig. 2.

\begin{figure}[tb]
\begin{center}
\includegraphics[width=8cm,clip]{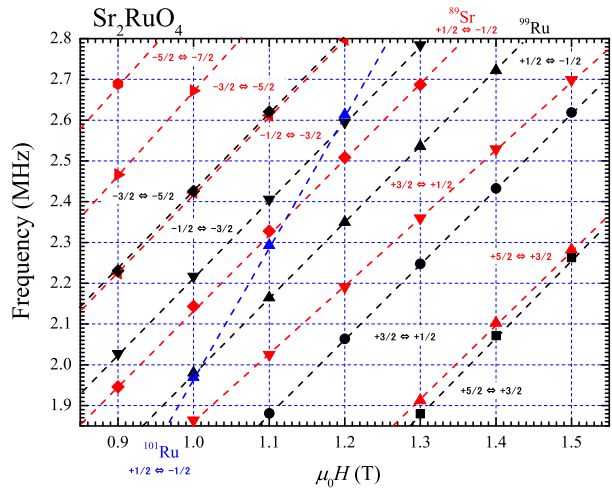}
\end{center}
\caption{(Color online) The NMR frequencies of each transition of $^{87}$Sr, $^{99}$Ru, and $^{101}$Ru in SRO under the basal plane. 
The frequencies are calculated by solving the secular equation with the Knight-shift values and quadrupole frequencies of SRO. 
The fractions indicate the eigenvalues of $I_z$.}
\label{FigS2}
\end{figure}

\begin{table}[htp]
\caption[]{Properties of Sr and Ru isotopes; the nuclear spin $I$, the gyromagnetic ratio $\gamma_n$, the nuclear quadrupolar moment $Q$, and the natural abundance N.A.} 
\label{table:Ru_parameter}
\vspace{5mm}
\begin{tabular}{l c c c c}\hline
               & $I$  &$\gamma_n/2\pi$(MHz/T) & $Q$(10$^{-24}$cm$^2$ )  & N.A.(\%) \\ \hline
$^{ 87}$Sr  & 9/2 & 1.845 & 0.3  & 7.0 \\
$^{ 99}$Ru  & 5/2 & 1.954 & 0.076 & 12.7 \\
$^{101}$Ru  & 5/2 & 2.193 & 0.44  & 17.1 \\ \hline
\end{tabular}
\end{table}

\section{Estimation of extra magnetization in the SC state}
In the following, we explain the origin of the extra magnetization in the SC state from the phenomenological discussion. 
This follows a part of Ref. \cite{Miyake14JPSJ}, in which the ground-state of SRO and order estimation of the extra magnetization are discussed, but we include simple diagrams to describe the origin. 
A physical reason for the extra magnetization in the SC state is rather simple if the equal-spin-pairing (ESP) is formed in SRO. 
Under the magnetic field, the density of states (DOS) of the normal state quasiparticles of up-spin, $N_{\uparrow}(\xi)$, and those of down-spin, $N_{\downarrow}(\xi)$, are different if the particle-hole symmetry is apparently broken, i.e., $N(\xi)$'s are not constant but have linear term in the energy $\xi$ measured from the chemical potential. 
Then, the free energy gains associated with Cooper pair condensation are different in general, resulting in a redistribution of up-spin and down-spin component so as to gain more total condensation energy. 
Therefore, depending on the sign of the linear term of $N(\xi)$, the extra magnetization change arises under the external magnetic field $H$.  

Similar mechanism was predicted by S. Takagi as a possible effect of discontinuity in the spin susceptibility at critical temperature $T_c$ where the superfluid $^3$He exhibits a second-order phase transition from the normal to the $A$ phase at $H = 0$ \cite{Takagi74PTP}. 
The paper by Takagi also predicts that in the $A_1$ phase, which emerges for $H \neq 0$ with only one spin species forming Cooper pairs, there exists an extra spin-polarization independent of $H$ other than the BCS-type contribution. 
On the other hand, it predicted that the extra magnetization disappears as the $A_1$ phase evolves into the $A$ phase where both up- and down-spin components are forming the Cooper pairs. 
This is because Takagifs theory did not take into account the redistribution of fermions with up- and down-spin components in the $A$ phase, while it took into account the migration of fermions in the normal down-spin band to the up-spin ESP state in the $A_1$ phase. 
(Note that the sign of magnetic moment of electron and $^3$He is opposite.) 
We reformulate Takagi's discussion and extend it to the ground state under the magnetic field $H$ \cite{Miyake14JPSJ}.  
To begin with, we assume the $\xi$ dependence of the DOS $N(\xi)$ without the magnetic field $H$ is given by 
\[
N(\xi) \approx N_{\rm F}\left(1+\frac{\alpha}{\epsilon_{\rm F}}\xi\right) ,  \hspace{3cm}                      {\rm  (S1)}
\]
\begin{figure}[tb]
\begin{center}
\includegraphics[width=8.7cm,clip]{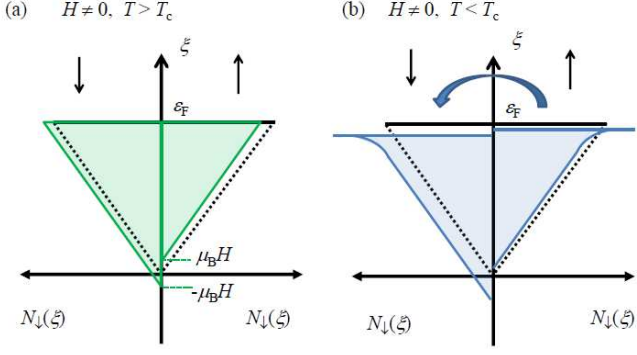}
\end{center}
\caption{(Color online)  (a) Schematic density of states (DOS) $N(\xi)$ vs energy $\xi$ of quasiparticles in the normal state. 
Dotted lines are DOS without magnetic field H. 
Green lines with $\mu_{\rm B}H$, and $-\mu_{\rm B}H$ shifts are DOS for up-spin and down-spin bands, respectively. 
The areas shown by shadows indicate the occupied states. 
A chemical potential shift due to the magnetic field is neglected as a negligible effect of the order of $O(\mu_B \mu_0 H/\epsilon_{\rm F})^2$. (b) DOS $N(\xi)$ vs $\xi$ of quasiparticles in the SC state. 
Due to the difference of DOS, the condensation energy for two spin states are different with each other, and thus some part of up-spin quasiparticles migrate to down-spin band so as to gain more total condensation energy.}
\label{FigS3}
\end{figure}
where $\alpha$  parameterizes steepness of the  slope of $N(\xi)$ around $\xi = 0$, and $\epsilon_{\rm F}$ is the effective Fermi energy. 
Then, the DOS of up spin, $N_{\uparrow}(\xi)$, and down-spin, $N_{\downarrow}(\xi)$, under the field $H$ are shifted as shown in Fig.~3 (a).
Here, we neglect the shift in the chemical potential of the order of $O(\mu_{\rm B} \mu_0H/\epsilon_{\rm F})^2$. 
Let us now discuss the consequence in the ground state based on a simple model. 
The difference of the condensation energy for down-spin and up-spin states in the ground state of ESP pairing is 
\[ 
\delta E_{\rm cond}=-\frac{1}{2}\left(N_{\downarrow}\Delta_{\downarrow}^2-N_{\uparrow}\Delta_{\uparrow}^2\right)\times\frac{1}{2}.  \hspace{1cm}            {\rm (S2)}
\]
With the use of the weak-coupling expression for the superconducting (SC) gap $\Delta'$s, $\Delta = \Omega \exp(-1/VN_{\rm F})$, as well as eq. (S1) for the DOS's, $\delta E_{\rm cond}$ is expressed as  
\begin{align*}
&\delta E_{\rm cond} \nonumber\\
&=
-\frac{1}{4}\Omega^2 \Biggl\{N_{\rm F} \left(1+\frac{\alpha}{\epsilon_{\rm F}}\mu_{\rm B}\mu_0H\right)\exp \left[-\frac{2}{VN_{\rm F}\left(1+\frac{\alpha}{\epsilon_{\rm F}}\mu_{\rm B}\mu_0 H \right) } \right]  \nonumber\\
& -N_{\rm F} \left(1-\frac{\alpha}{\epsilon_{\rm F}}\mu_{\rm B}\mu_0H\right)\exp \left[-\frac{2}{VN_{\rm F}\left(1-\frac{\alpha}{\epsilon_{\rm F}}\mu_{\rm B}\mu_0 H \right) } \right] \Biggr\}  \nonumber, \\
&                    \hspace{70mm}  {\rm (S3)}
\end{align*}
where we used the relations of $N_{\uparrow} = N_{\rm F}(1+\alpha\mu_{\rm B}\mu_0H/\epsilon_{\rm F})$ and $N_{\downarrow} = N_{\rm F}(1-\alpha\mu_{\rm B}\mu_0H/\epsilon_{\rm F})$. 
Therefore, up to the linear term in $H$, the $\delta E_{\rm cond}$ is given as
 \[
\delta E_{\rm cond}=-\frac{\Delta^2}{2}N_{\rm F}\frac{\alpha}{\epsilon_{\rm F}}\mu_{\rm B}\left(1+\frac{2}{VN_{\rm F}}\right)\mu_0H .    \hspace{5mm}   {\rm  (S4)}
\]
If $\alpha > 0$ as shown in Fig.~3 (b), under the magnetic field $H$, some part of the up-spin quasiparticles migrate to the down-spin band so as to gain more total condensation energy, enhancing the magnetization $m$ to $m+\delta m$. 
This extra magnetization $\delta m$ is determined by the condition that the energy gain compensates the energy loss $\delta E_{\rm mag}$ due to the extra magnetization process. 
The $\delta E_{\rm mag}$ is given as 
\begin{eqnarray*}
\delta E_{\rm mag} &=& \int_m^{m + \delta m}\mu_0HdM=\frac{1}{2\chi}(m+\delta m)^2-\frac{1}{2\chi}m^2 \\
 &\approx& \frac{m \delta m}{\chi} \approx \mu_0 H \delta m, \hspace{10mm}   {\rm (S5)}
\end{eqnarray*}                  ,          
where $\chi$ represents the uniform susceptibility in the normal state. 
By equating magnitudes of the right hand sides of eqs. (S4) and (S5), we obtain
\[
 \delta m \cong -\frac{\Delta^2}{2}N_{\rm F}\frac{\alpha}{\epsilon_{\rm F}}\mu_{\rm B}\left(1+\frac{2}{VN_{\rm F}}\right).    \hspace{5mm}   {\rm (S6)}
\]
Equation (S6) is also derived on much more deliberate calculations in Ref. \cite{Miyake14JPSJ}.
The magnetization $m_n$ in the normal state under the magnetic field $H$ is given by $m_n \cong 2\mu_B^2N_{\rm F}\mu_0H/(1+F_0^a)$, where the Landau parameter for correction of the susceptibility. 
Therefore, the ratio of $\delta m$ and $m_n$ is given by 
\begin{equation*}
\frac{\delta m}{m_n} \cong \frac{\alpha \Delta^2}{4\mu_{\rm B} \epsilon_{\rm F} \mu_0 H}\left(1+\frac{2}{VN_{\rm F}}\right)\left(1+F_0^a\right). \hspace{3mm}  {\rm (S7)} 
\end{equation*}                      
Note that this ratio vanishes as long as the particle-hole symmetry is retained ($\alpha$ = 0).
Here we give a rough order estimation for $\delta m/m$ in SRO.  The effective Fermi energy of the quasiparticles $\epsilon_F$ is estimated as $\epsilon_{\rm F} \approx 3.8 \times 10^3$ K \cite{Miyake10JPSJ}. 
Assuming $\Omega \sim \epsilon_{\rm F}$, the coupling constant $VN_{\rm F}$ is estimated as $1/VN_{\rm F} \sim 7$. 
The SC gap at $T$ = 0 K is estimated by using the BCS relation, $\Delta \cong 1.7 T_c \approx 2.6$ K. 
The magnetic field $\mu_0H$ = 1.0 T is equivalent to 0.67 K. 
The Landau parameter is $F_0^a \sim  -1/2$\cite{Mackenzie03RMP,Miyake10JPSJ}.  
Then the ratio $\delta m/m_n$ at $T$ = 0 K is estimated to be 
\[
 \frac{\delta m}{m_n} \approx 5.5 \times 10^{-3} \alpha.  \hspace{5mm}   {\rm (S8)}
\]

Since there exists a van Hove singularity in the DOS of the $\gamma$ band, the parameter $\alpha$ is estimated to be $\alpha \sim 4$\cite{Bergemann03AdvPhys}, which is much larger than 1/2, the value of free fermions. Thus the ratio $\delta m/m_n$ at $T$ = 0 K can be a few percent, in good agreement with the Knight shift measurement reported in this paper.

\end{document}